\documentstyle[aps,epsfig,prl]{revtex}

\newcommand{\Rpp}
             {$R^0\rightarrow\tilde{\gamma}\rho,\; \rho\rightarrow\pi^+\pi^-$}
\newcommand{\rkr}{$R^0/K_L$ flux ratio }

\begin{document}
\draft
\title{Search for Light Gluinos via the Spontaneous Appearance of 
$\mathbf{\pi^+\pi^-}$ Pairs with an 800~GeV/c Proton Beam at Fermilab}

\author{
J.~Adams$^{11}$,
A.~Alavi-Harati$^{12}$,
I.F.~Albuquerque$^{10}$,
T.~Alexopoulos$^{12}$,
M.~Arenton$^{11}$,
K.~Arisaka$^2$,
S.~Averitte$^{10}$,
A.R.~Barker$^5$,
L.~Bellantoni$^7$,
A.~Bellavance$^9$,
J.~Belz$^{10}$,
R.~Ben-David$^7$,
E.~Blucher$^4$,
G.J.~Bock$^7$,
C.~Bown$^4$,
J.D.~Bricker$^{10}$
S.~Bright$^4$,
E.~Cheu$^1$,
S.~Childress$^7$,
R.~Coleman$^7$,
M.D.~Corcoran$^9$,
G.~Corti$^{11}$,
B.~Cox$^{11}$,
M.B.~Crisler$^7$,
A.R.~Erwin$^{12}$,
R.~Ford$^7$,
G.~Graham$^4$,
J.~Graham$^4$,
K.~Hagan$^{11}$,
E.~Halkiadakis$^{10}$,
K.~Hanagaki$^8$,
S.~Hidaka$^8$,
V.~Jejer$^{11}$,
J.~Jennings$^2$,
D.A.~Jensen$^7$,
P.~Johnson$^7$,
R.~Kessler$^4$,
H.G.E.~Kobrak$^{3}$,
J.~LaDue$^5$,
A.~Lath$^{10}$,
A.~Ledovskoy$^{11}$,
A.P.~McManus$^{11}$,
P.~Mikelsons$^5$,
E.~Monnier$^4$\cite{p2},
T.~Nakaya$^7$,
U.~Nauenberg$^5$,
K.S.~Nelson$^{11}$,
H.~Nguyen$^7$,
V.~O'Dell$^7$,
M.~Pang$^7$,
R.~Pordes$^7$,
V.~Prasad$^4$,
C.~Qiao$^4$,
B.~Quinn$^4$,
E.~Ramberg$^7$,
R.E.~Ray$^7$,
A.~Ronzhin$^7$,
A.~Roodman$^4$,
M.~Sadamoto$^8$,
S.~Schnetzer$^{10}$,
K.~Senyo$^8$,
P.~Shanahan$^7$,
P.~Shawhan$^4$,
W.~Slater$^2$,
N.~Solomey$^4$,
S.V.~Somalwar$^{10}$\cite{p1}, 
R.L.~Stone$^{10}$,
E.C.~Swallow$^{6,4}$,
R.A.~Swanson$^{3}$,
R.J.~Tesarek$^{10}$,
G.B.~Thomson$^{10}$,
R.~Tschirhart$^7$,
Y.W.~Wah$^4$,
H.B.~White$^7$,
J.~Whitmore$^7$,
B.~Winstein$^4$,
R.~Winston$^4$,
J.-Y.~Wu$^5$,
T.~Yamanaka$^8$,
E.D.~Zimmerman$^4$
}

\author{(KTeV Collaboration)\vspace{.5cm}}

\address{
$^1$ University of Arizona, Tucson, Arizona 85721 \\
$^2$ University of California at Los Angeles, Los Angeles, California 90095 \\
$^{3}$ University of California at San Diego, La Jolla, California 92093 \\
$^4$ The Enrico Fermi Institute, The University of Chicago,
Chicago, Illinois 60637 \\
$^5$ University of Colorado, Boulder, Colorado 80309 \\
$^6$ Elmhurst College, Elmhurst, Illinois 60126\\
$^7$ Fermi National Accelerator Laboratory, Batavia, Illinois 60510 \\
$^8$ Osaka University, Toyonaka, Osaka 560 Japan \\
$^9$ Rice University, Houston, Texas 77005 \\
$^{10}$ Rutgers University, Piscataway, New Jersey 08855 \\
$^{11}$ University of Virginia, Charlottesville, Virginia 22901 \\
$^{12}$ University of Wisconsin, Madison, Wisconsin 53706 \\
}

\maketitle

\vspace{-5.2in}
\begin{flushright}
RUTGERS-97-26\\
\end{flushright}
\vspace{5.0in}
\begin{center}(Rutgers-97-26, Fermilab-Pub-97-320-E, hep-ex/9709028, 
               To Appear in Phys. Rev. Lett.)\end{center}

\begin{abstract}
We searched for the 
appearance of $\pi^+\pi^-$ pairs 
with invariant mass  $\geq$648~MeV/c$^2$ in a neutral beam.
Such an observation
could signify the decay of a long-lived light neutral particle.
We find no evidence for this decay. 
%
%
Our null result severely constrains the
existence of an $R^0$ hadron, which is the lightest bound state of a
gluon and a light gluino $(g\tilde{g})$, and thereby also the possibility 
of a light gluino.  Depending on the photino
mass, we exclude the $R^0$ in the mass and lifetime ranges 
of 1.2 -- 4.6~GeV/c$^2$ and 
$2\times10^{-10}$--$7\times10^{-4}$~s, respectively.
\end{abstract}




This Letter is motivated by recent discussions~\cite{ref:FAR1} 
of the possible existence of long-lived hadrons that contain light gluinos. 
In some theories where the breaking of
supersymmetry is communicated to ordinary particles 
by the exchange of very heavy states,
gauginos have small tree-level masses and are light compared to the squarks.
The gluino($\tilde{g}$) and photino($\tilde{\gamma}$) masses are 
expected to be $\lesssim1.0$~GeV/c$^2$.
The gluons ($g$), gluinos, and quarks can form 
bound states, the lightest of which is a spin-1/2 $g\tilde{g}$
combination called the $R^0$. The approximate degeneracy of $R^0$ with
the 0$^{++}$ gluonia~\cite{ref:FAR1} suggests that the 
$R^0$ mass could be of the order of 1.3--2.2~GeV/c$^2$.
The stable photino in this theory is a cold dark matter 
candidate~\cite{ref:FAR3}. Estimates from 
particle physics~\cite{ref:FAR1} and cosmology~\cite{ref:FAR3} place the
$R^0$ lifetime in the $10^{-10}$--$10^{-5}$~s range.
The $R^0$ decays into a $\tilde{\gamma}$ and hadrons.
Because of the approximate $C$ invariance of supersymmetric QCD, the $R^0$ 
decay into $\rho\tilde{\gamma}$ is expected to be the dominant 
decay mode~\cite{ref:FAR1};
depending on the extent of $C$ violation, the $R^0$ may 
also decay into  $\pi^0\tilde{\gamma}$ or $\eta\tilde{\gamma}$.
The existence of a light gluino can have an impact on the running of
the strong coupling constant $\alpha_s$.
A phenomenological analysis of the perturbative running of 
$\alpha_s$\cite{ref:FOD}, in conjunction with the multijet analysis at 
LEP\cite{ref:ALEPH} has claimed an indirect exclusion of the light gluino 
scenario.  However, a debate over the extent of this 
exclusion\cite{ref:GFLT} underlines the necessity of a direct search
for hadrons containing light gluinos.

The $R^0$'s can be produced in $pN$ collisions by processes such as 
quark-antiquark annihilation, gluon fusion, etc. Since squarks need
not be involved, the $R^0$ production is
in the realm of conventional QCD. Dawson, Eichten, and 
Quigg\cite{ref:QUG1} have calculated cross sections for gluino production
in tree approximation.  Their calculations suggest an order-of-magnitude 
cross section estimate\cite{ref:QUG2}
of $\sim10~\mu$b per nucleon for the production of an $R^0$ with
2~GeV/c$^2$ mass in 800~GeV/c $pN$ collisions. 
This cross section corresponds to
$\sim10^{-3}$ $R^0$ to $K_L$ flux ratio in our experiment, as explained later.
Assuming a $10^{-8}$~s lifetime and 
a 100\% branching ratio for the decay of this $R^0$ into 
$\tilde{\gamma}\rho,\; \rho\rightarrow\pi^+\pi^-$, the
KTeV experiment~\cite{ref:CDR} has a per spill sensitivity of
approximately $10^{-4}$ in terms of the $R^0$ to $K_L$ flux ratio with
$3.5 \times 10^{12}$ 800~GeV/c protons delivered in a nominal spill.
The results presented here are from the delivery of
$1.9 \times 10^{15}$ protons.

\begin{figure}
 \begin{center}
   \parbox{3in}{\epsfxsize=3in\epsffile
                                          {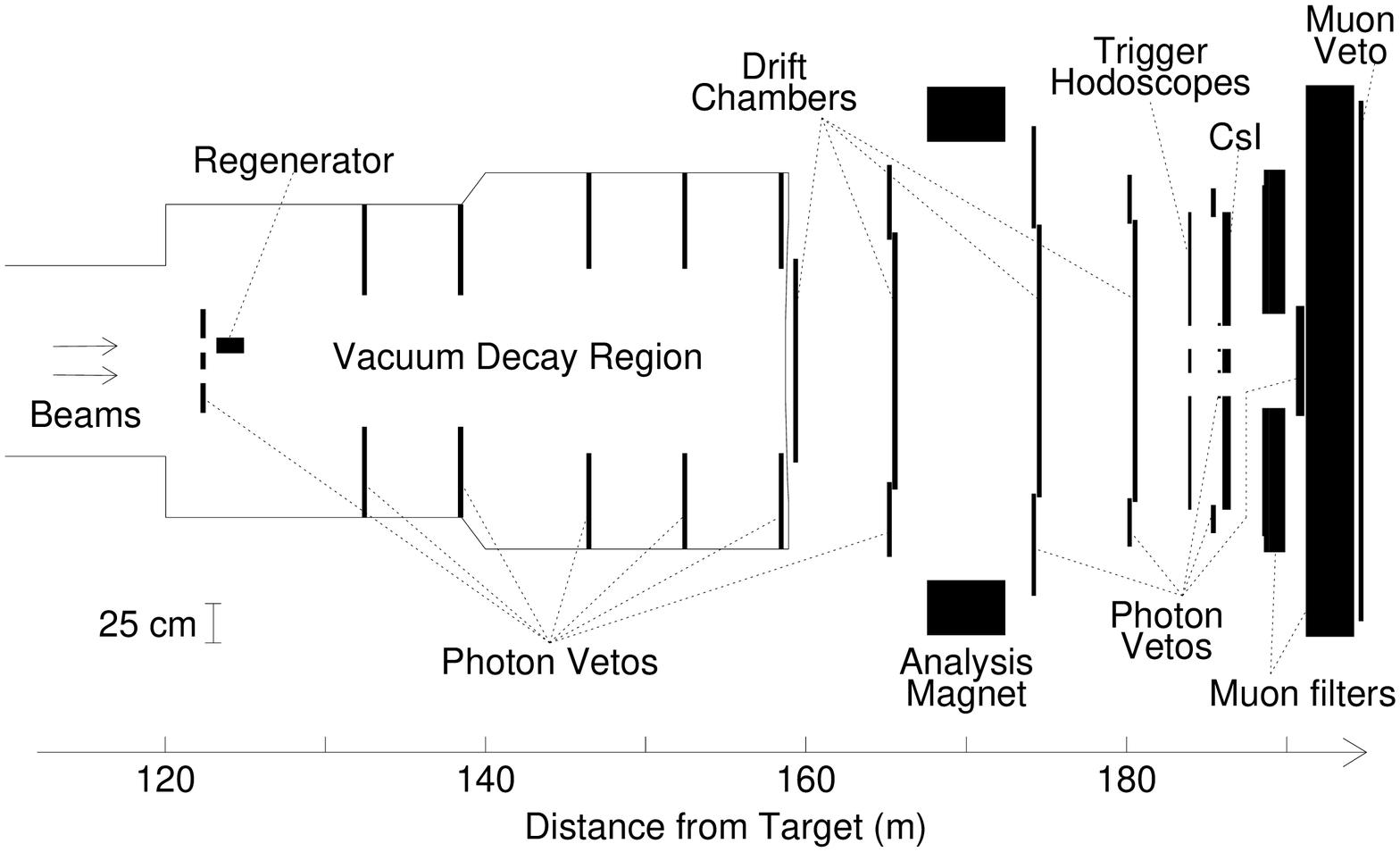}}
 \end{center}
  \caption{KTeV detector configuration for this measurement.
           Note the highly compressed scale along the beam direction.
          }
  \label{fig:det}
\end{figure}

In the KTeV experiment (figure~\ref{fig:det}), protons are incident 
on a 30~cm
long beryllium oxide target at a vertical angle of 4.8~mr with
respect to the neutral beam channel. The interaction products are filtered
through a 50.8~cm beryllium absorber and a 7.6~cm lead absorber.
Two neutral beams, each 0.25~$\mu$str in solid angle, emerge
after collimation and sweeping. One of the beams passes through 
an active regenerator, but the decays from this beam are not 
used in this analysis. The beam transport and decays took place in an 
evacuated region with a vacuum of $0.5$--$1.0 \times 10^{-4}$~torr.

The most crucial detector element for this analysis, the charged
spectrometer, consists of four planar drift
chambers, two on either side of an analyzing magnet.
Each chamber measures positions in two orthogonal views.
Each view consists 
of two planes of wires, with cells arranged in a hexagonal geometry. 
Each chamber has approximately 100~$\mu$m single-hit position resolution 
per plane. The 
spectrometer magnet has a 2~m by 1.7~m fiducial region over which
a transverse momentum impulse of 411~MeV/c is imparted to the charged
particles. A set of helium bags integrated
into the spectrometer system minimizes multiple scattering.
The invariant mass resolution for
the decay $K_L \rightarrow \pi^+\pi^-$ is better than 2~MeV/c$^2$.

A pure CsI electromagnetic calorimeter of dimension 1.9~m by 1.9~m is used 
to reconstruct photon and electron energies to better than 1\% precision.
The calorimeter is used to match the orthogonal
track views and to reject background from $K_{e3}$ decays.  A set of 12 
photon vetos provides hermetic photon coverage up to angles of 100~mr.
A counter bank (muon veto) located at the
downstream end of the detector is used to reject $K_{\mu3}$ decays.
The event trigger is initiated by signals from two scintillator hodoscopes
located downstream of the spectrometer.  The primary trigger 
requires two hits in these counters consistent with two oppositely charged
tracks, at least one hit in each of the two upstream drift chambers, and
lack of hits in the muon veto bank.  Next, a set of fast trigger processors 
requires two straight tracks in the non-bend view from two hits 
in each drift chamber.

\begin{figure}
 \begin{center}
   \parbox{3in}{\epsfxsize=3in\epsffile
                                               {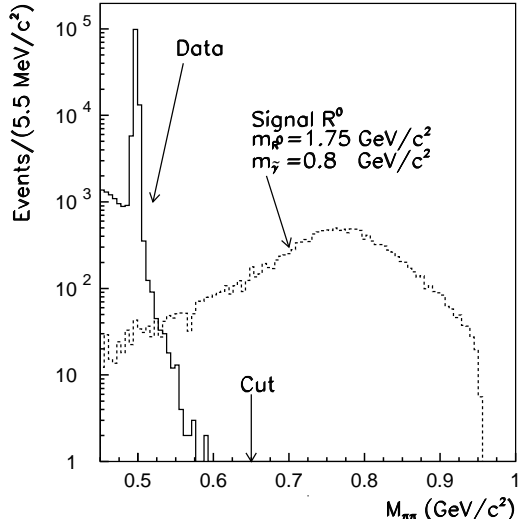}}
 \end{center}
  \caption{$m_{\pi^+\pi^-}$ distribution for the data (solid)
           and $R^0$ signal Monte Carlo (dashed, arbitrary scale).  The peak 
           at 
           500~MeV/c$^2$ corresponds to $K_L \rightarrow \pi^+\pi^-$ decays.
           The sharp cutoff
           at 0.95 GeV/c$^2$ (=$m_{R^0}-m_{\tilde{\gamma}}$) is due to the
           kinematic limit for the $R^0$ shown.
          }
  \label{fig:data}
\end{figure}

	A combination of online and offline cuts in the $\pi^+\pi^-$ 
analysis required two oppositely charged tracks matched to clusters 
in the calorimeter to within 4.6~cm.  Individual track momenta were
required to be more than 8~GeV/c, and the scalar sum of two track momenta was
required to be between 30~GeV/c and 160~GeV/c.  A longitudinal vertex
position between 126~m and 155~m downstream of the target defined the fiducial
region.
Electrons were rejected by requiring the energy deposited in the
calorimeter to be less than 80\% of the particle momentum, as measured in 
the spectrometer. The signals in various veto devices 
were required to be no more than those due to accidental activity.
The two tracks were required to project within the active
fiducial area of the muon banks.
After making all other cuts, the momentum ratio of 
the two tracks was required to be between 0.2~and~5, 
which curtailed the high-side tail
of the $\pi^+\pi^-$ mass distribution by moving its end point from
$\sim$640~MeV/c$^2$ to $\sim$600~MeV/c$^2$.
Figure~\ref{fig:data} shows the $m_{\pi^+\pi^-}$
distribution for the events surviving all cuts.  Note the peak 
corresponding to the
$K_L \rightarrow \pi^+\pi^-$ candidates and 
the rapidly falling background to the right of the kaon peak.
The figure also shows the $m_{\pi^+\pi^-}$ 
distribution for an $R^0$ with mass $m_{R^0}$=1.75~GeV/c$^2$ and photino mass
$m_{\tilde{\gamma}}$=0.8~GeV/c$^2$.
There are no $R^0$ candidates 
in the signal window between 648~MeV/c$^2$ (i.e. 150~MeV/c$^2$ above the 
kaon mass) and 1.0~GeV/c$^2$. 
For the simulated $R^0$ shown, 91\% of the decays are contained in the signal
window.

The absence of  signal in the data can be expressed in terms of 
upper limits on the $R^0$ flux. We define the
$R^0/K_L$ flux ratio to be the ratio of the number of $R^0$'s to the number 
of $K_L$'s exiting the 
beam absorbers, calculated with the assumptions that the \Rpp branching 
ratio is 100\% and that the photino does not interact significantly 
in the detector material.
For the $R^0$ spectrum shape, we use the invariant cross section 
for $\Lambda$ 
production in $pBe$ interactions\cite{ref:PON}.  We make this choice because
the $R^0$ mass is in the proximity of the $\Lambda$-$\Xi^0$-$D^0$ mass range.
For our experiment, the function
$(1-x_F)^a$exp$(-bp_{\perp}^2)$ with $a=1.0$ and
$b=2.3$(GeV/c)$^{-2}$ is a good approximation for the cross section shapes 
given in \cite{ref:PON} for both $\Lambda$ and $\Xi^0$.
The limits are not very sensitive to the spectrum shape; if we use the
values $a=6.1$ and $b=1.08$(GeV/c)$^{-2}$ which are applicable to $D^0$
production\cite{ref:ALV}, the limits change by $\sim$5-10\% for $R^0$ 
lifetime of 10$^{-8}$~s.

\begin{figure}
 \begin{center}
   \parbox{3.0in}{\epsfxsize=3.0in\epsffile
                                                {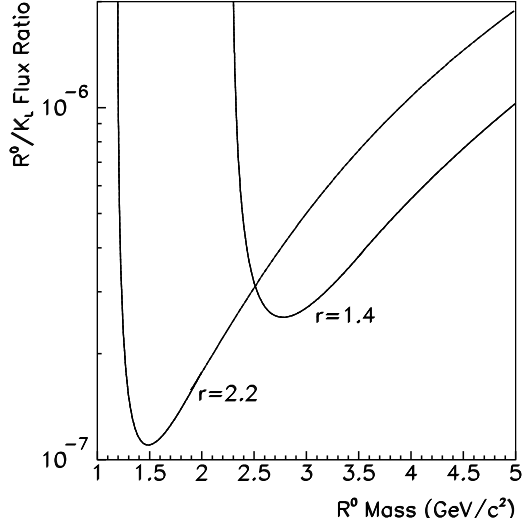}}
 \end{center}
  \caption{Upper limits with 90\% confidence level on the \rkr
           as a function of $R^0$ mass for two different values of the
           $R^0$-photino mass ratio $r$. ($R^0$ lifetime =  $10^{-8}$s)
          }
  \label{fig:mass}
\end{figure}

The $K_L$ flux was determined using a total of 116,552
$K_L\rightarrow\pi^+\pi^-$ candidates
with two-body transverse momentum squared ($p_{t}^2$)
less than 250~MeV$^2$/$c^2$ and $m_{\pi^+\pi^-}$
within 10~MeV/c$^2$ of the $K_L$ mass.
The detector acceptance for the $K_L \rightarrow \pi^+\pi^-$ decays
was calculated using
a Monte Carlo simulation of the beam and the detector.  
We calculate a
\begin{figure}
 \begin{center}
   \parbox{3.0in}{\epsfxsize=3.0in\epsffile
                                                   {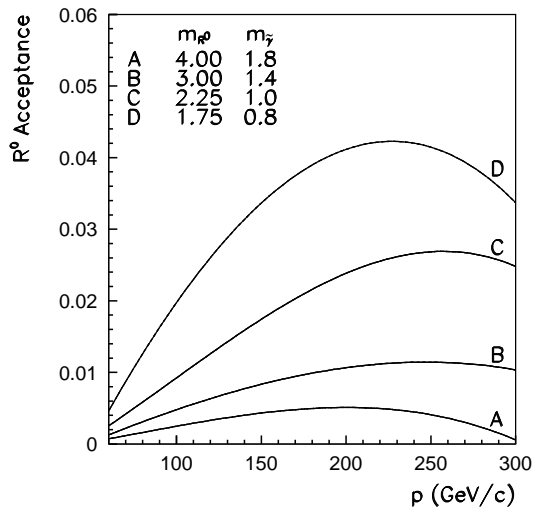}}
 \end{center}
  \caption{Geometrical acceptance of the detector in the fiducial region
           for various $R^0$ masses as a function of the $R^0$ momentum.
           $m_{R^0}$ and $m_{\tilde{\gamma}}$ are in GeV/c$^2$.
          ($R^0$ lifetime =  $10^{-8}$s)
            }
  \label{fig:acc}
\end{figure}
\noindent
total of $1.33 \times 10^{10}$ $K_L$'s of all
energies exiting the absorbers.
To calculate the detector acceptance for $R^0$'s, the \Rpp decays
were simulated for various values of $m_{R^0}$ and $m_{\tilde{\gamma}}$ 
assuming isotropic angular distributions of the decay products in the
center of mass frame. Figure~\ref{fig:acc} 
shows the geometric acceptance of the detector in the fiducial region
for various $R^0$ masses as a function of the $R^0$ momentum. 
Figure~\ref{fig:mass} shows the 90\% confidence level 
upper limits on the \rkr
as a function of $m_{R^0}$ with a 
$10^{-8}$~s $R^0$ lifetime for two values of 
the mass ratio $r=m_{R^0}/m_{\tilde{\gamma}}$. 
The sharp loss of sensitivity for small $m_{R^0}$ for a given $r$ is because
an $R^0$ with mass less than $0.648r/(r-1)$~GeV/c$^2$ can not
produce a $\tilde{\gamma}$ with mass  $m_{R^0}/r$ together with
a $\pi^+\pi^-$ pair having
invariant mass greater than 648~MeV/c$^2$.

\begin{figure}
 \begin{center}
   \parbox{3.0in}{\epsfxsize=3.0in\epsffile
                                                   {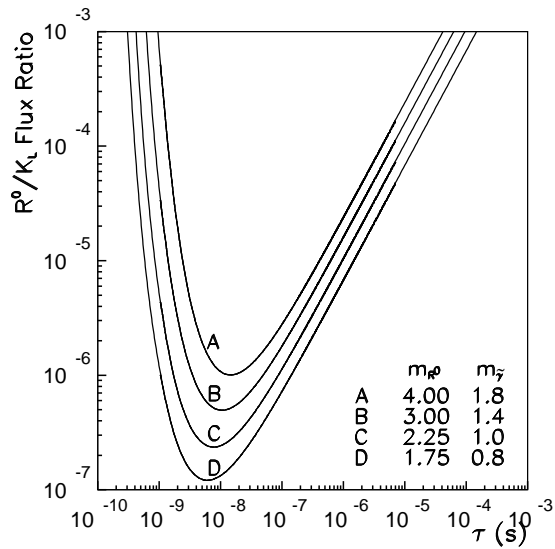}}
 \end{center}
  \caption{Upper limits with 90\% confidence level on the \rkr
           as a function of $R^0$ lifetime. $m_{R^0}$ and 
           $m_{\tilde{\gamma}}$ are in GeV/c$^2$.($r$=2.2)
          }
  \label{fig:time}
\end{figure}

\begin{figure}
 \begin{center}
   \parbox{3.0in}{\epsfxsize=3.0in\epsffile
                                             {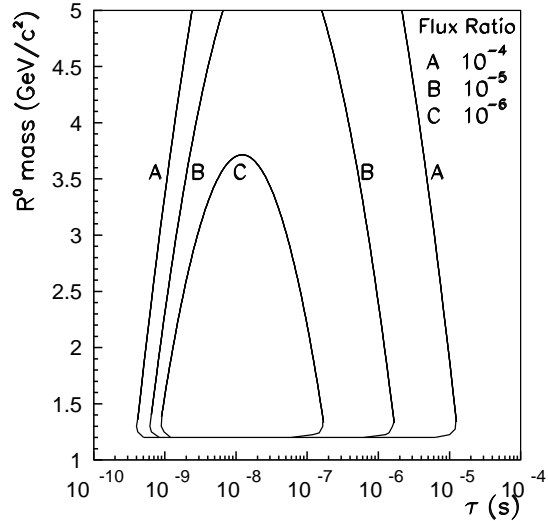}}
 \end{center}
  \caption{Upper limits with 90\% confidence level on the \rkr
           for various $R^0$ masses and lifetimes.($r$ =2.2)
          }
  \label{fig:c1}
\end{figure}

\begin{figure}
 \begin{center}
   \parbox{3.0in}{\epsfxsize=3.0in\epsffile
                                            {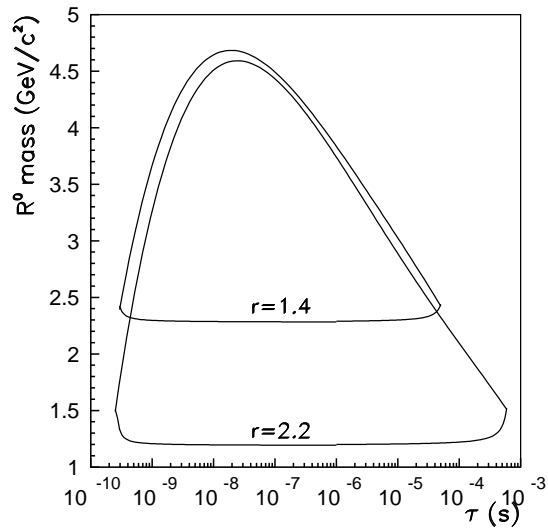}}
 \end{center}
  \caption{$R^0$ mass-lifetime regions excluded by this analysis at
           90\% confidence level for two different values of the
           $R^0$-photino mass ratio $r$.}
  \label{fig:c2}
\end{figure}

Figure~\ref{fig:time} shows the upper limits on \rkr
as a function of $R^0$ lifetime with different masses for 
the same $r$(=2.2).
Figure~\ref{fig:c1} shows the 90\% confidence level upper limit contours
for various $R^0$ masses and lifetimes. 
These upper limits can also be expressed in terms of limits on 
the invariant cross section $Ed^3\sigma/dp^3$ times the branching ratio 
using the conversion factor of 
$2.1\times 10^{-36}$cm$^2$/(GeV$^2$/c$^3$) at $x_F = 0.2$
per $1 \times 10^{-7}$ \rkr. 
%
Our limits are a significant improvement over the previous 
(indirect) search by Bernstein {\it et al.}~\cite{ref:BER}.
Note that this conversion factor and the upper limit on $Ed^3\sigma/dp^3$
reported by Bernstein {\it et al.} do not take
into account the $R^0$ absorption in the target and
absorbers. The $R^0N$ cross section is expected~\cite{ref:FAR1,ref:FAR4} 
to be in the range 1/10 to 1 times the $NN$ 
cross section, which places the $R^0$ absorption factor in the
1.3 to 10.2 range.
The upper limits on the \rkr constrain the light gluino
scenario assuming a specific model for the $R^0$ production.
The perturbative QCD calculations\cite{ref:QUG1,ref:QUG2} suggest
$\sigma(R^0)\simeq860e^{-2.2m_{R^0}}~\mu$b in
800~GeV/c $pN$ collisions, where $m_{R^0}$ is in the units of GeV/c$^2$. This
estimate is approximately consistent with the heavy flavor production
cross sections. 
For our experiment, this $\sigma(R^0)$ implies that the \rkr for an $R^0$
of mass $m_{R^0}$ is expected to be $9.2\times10^{-2}e^{-2.2m_{R^0}}$,
assuming a factor of 10.2 for the $R^0$ absorption.
The mass-lifetime region for which this \rkr expectation
exceeds the measured upper limits is taken to be ruled out.
For example, the expected \rkr for an 
$R^0$ with 2~GeV/c$^2$ mass is $1.1\times10^{-3}$. Since this expectation
exceeds our upper limits in the lifetime range of
$3.4\times10^{-10}$ -- $1.3\times10^{-4}$~s, we rule 
out $R^0$'s with 2~GeV/c$^2$ mass in this lifetime range with 90\% 
confidence level.  
Figure~\ref{fig:c2} shows the excluded regions obtained in this fashion
for two different values of the mass 
\noindent
ratio $r$. Apart 
from the low mass cutoff
for $m_{R^0}\lesssim0.648r/(r-1)$~GeV/c$^2$, the contour 
shapes in 
figure~\ref{fig:c2} are fairly insensitive to the value of mass ratio $r$.
The contours for the cosmologically interesting region correspond to
$r\lesssim 1.8$~\cite{ref:FAR3}.

In summary, our search 
places stringent upper limits on the \rkr. For example, the 90\% confidence
level upper limit on the \rkr for
an $R^0$ with mass 1.75~GeV/c$^2$, lifetime $6\times10^{-9}$~s,
and mass ratio $r$=2.2 is  $1.2\times10^{-7}$.
As figure~\ref{fig:c2} shows, these stringent limits rule out
the production of $R^0$ particle in a wide range of masses
($\sim$1.2 -- 4.6~GeV/c$^2$) and lifetimes
($\sim3\times10^{-10}$ -- $7\times10^{-4}$~s).
A search\cite{ref:761} for light supersymmetric {\em baryons} 
($uud\tilde{g}$ and $uds\tilde{g}$) in the 
mass range 1.7--2.5~GeV/c$^2$ has also reported null results.

We thank G.R.~Farrar for suggesting this search, W.~Molzon, C.~Quigg, and
B.~Schwingenheuer for 
helpful conversations, and Fermilab staff for their dedication.
This work was supported by the NSF, DOE, and the US-Japan 
Foundation. A.R.B.,E.B. and S.V.S. acknowledge support from the NYI program 
of the NSF, A.R.B. and E.B. from the A.P.~Sloan Foundation, E.B. from
the OJI program of the DOE, and K.H., T.N., and M.S. from the JSPS.


\begin{references}

\bibitem [*]{p1}        To whom correspondence should be addressed.\\
                        Electronic address: somalwar@physics.rutgers.edu

\bibitem [\dagger]{p2}  On leave from C.P.P. Marseille/C.N.R.S., France.

\bibitem{ref:FAR1}	G.R. Farrar,
                        Phys. Rev. Lett. {\bf76}, 4111 (1996);
			G.R.~Farrar,
                        Phys. Rev. D {\bf51}, 3904 (1995).

\bibitem{ref:FAR3}	D.J.H.~Chung, G.R.~Farrar, and E.W.~Kolb, 
                        Report No. astro-ph/9703145;
			G.R.~Farrar and E.W.~Kolb,
                        Phys. Rev. D {\bf53}, 2990 (1996).

\bibitem{ref:FOD}	F. Csikor and Z. Fodor,
                        Phys. Rev. Lett. {\bf78}, 4335 (1997).

\bibitem{ref:ALEPH}	R. Barate {\it et al.},
                        Report No. CERN-PPE-97-002.

\bibitem{ref:GFLT}	G.R. Farrar, in Proceedings of the Rencontres
                        de la Vallee d'Aoste, La Thuille, March 1997,
                        Report No. Rutgers-97-22
                        (to be published).

\bibitem{ref:QUG1}	S. Dawson, E. Eichten, and C. Quigg,
                        Phys. Rev. D {\bf31}, 1581 (1985).

\bibitem{ref:QUG2}	C.~Quigg (private communication).

\bibitem{ref:CDR}	K. Arisaka {\it et al.},
                        Report No. FERMILAB-580-1992;
                	L.K.~Gibbons {\it et al.},
                        Phys. Rev. D {\bf 55} 6625 (1997).

\bibitem{ref:PON}	L.G. Pondrom,
                        Phys. Rep. {\bf122}, 57 (1985).

\bibitem{ref:ALV}	G.A. Alves {\it et al.},
                        Phys. Rev. Lett. {\bf 77}, 2392 (1996);
                        K.~Kodama {\it et al.},
                        Phys. Lett. B {\bf 263}, 573 (1991).

\bibitem{ref:BER}	R.H. Bernstein {\it et al.},
                        Phys. Rev. D {\bf37}, 3103 (1988).

\bibitem{ref:FAR4}	G.R. Farrar (private communication).

\bibitem{ref:761}	I.F. Albuquerque {\it et al.},
                        Phys. Rev. Lett. {\bf78}, 3252 (1997).

\end{references}
\end{document}